\documentclass[a4paper,11pt]{article}
\begin{document}
\title{Explicitly correlated Helium wave function and hyperspherical coordinates}
\author{R.Habrovsk\'y\\
Slovak Hydrometeorological Institute, Jes\'eniova 17, SK-833 15 Bratislava,\\
Slovak Republic\\
email: richard.habrovsky@shmu.sk}

\maketitle

\begin{abstract}

Wave functions of a new functional kind have been proposed 
for Helium-like atoms in this work .
These functions explicitly depend on interelectronic and  
hyperspherical coordinates.
The best ground state energy for the Helium atom 
$ -2.903724376677 a.u.$ 
has been calculated with variational method  with basis set of  simple 
functions with a single exponential parameter. 
To the author's knowledge, this is the best result 
with use of hyperspherical coordinates so far.
Comparable result has been obtained for the hydrogen anion. 
For Helium atom, our best wave functions matched 
the Kato cusp conditions within the accuracy below $6.10^{-4} $.
An important feature of proposed wave functions is the inclusion 
of negative powers of $R=\sqrt(r^{2}_{1}+r^{2}_{2})$ in combination 
with positive powers of $r_{12}$ into the wave function. 
We showed that this is necessary condition for proposed wave function to be
a formal solution of Schr\"odinger equation.
\end{abstract}

\section{Introduction}
\label{intro}
It is well known from the time of Hylleraas \cite{RefHyll1,RefHyll2}, that one of the necessary conditions of a relevant
convergence towards the exact nonrelativistic ground state Helium energy is the inclusion of
$r_{12}$ terms into the wave function.  
Until now many different methods have been suggested 
where the $r_{12}$ function is used to construct the wave function. 
These methods could be divided to
variational \cite{RefHyll1,RefHyll2,RefKin,RefThak,RefFran,RefFreund,RefGoldm,RefKor,RefDrak1,RefDrak2,RefSim,RefSchwa}, correlated-function hyperspherical-harmonic (HH) methods \cite{RefHaft,RefWan,RefKri} 
(nice overview of HH methods is in the paper of Krivec \cite{RefKri}) and ICI method \cite{RefNak1,RefNak2,RefNak3}. 
Generally we can say that variational methods converge
	to globally optimized solutions while hyperspherical-harmonic methods converge
pointwise. However it is known (Bartlett et. all \cite{RefBar1}), that wave function created from only the Slater functions and  powers of $r_{1}$,$r_{2}$ and $r_{12}$ is not the exact one. 
It was Bartlett \cite{RefBar2} and Fock \cite{RefFock} who first pointed out that exact wave function must depend on coordinate
 $R=\sqrt{r_{1}^{2}+r_{2}^{2}}$. Fock's wave function depends on the powers of $R^{2}$ multiplied by powers of the logarithmic term $\ln{R^{2}}$. 
Due to the mathematical difficulty of direct implementation of Fock's approach 
it was probably first implemented (in some approximate way) in work of Frankowski 
and Pekeris \cite{RefFran} in 1966. Their results were improved by Freund, Huxtable and Morgan \cite{RefFreund}, each group 
 used wave functions that were some mixture of original Hylleraas coordinates and powers of the 
 logarithmic term of the coordinate $s=r_{1}+r_{2}$. 

Our effort is concentrated on finding a competitive approach with use of simple functions as coordinate system that could be generalized 
to more than two electron atoms. We proposed the approach where the wave function depends besides
the powers of $r_{12}$ on the hyperradial coordinate
$R=\sqrt{r_{1}^{2}+r_{2}^{2}}$ and on the coordinate 
$t=(r_{2}^{2}-r_{1}^{2})/(r_{2}^{2}+r_{1}^{2})$.
The function $t$ has only one shortage, it is not well defined at the point where $r_{1}=0$ and $r_{2}=0$, so it 
has not derivative in this point. But in the rest area it is well defined, so it is not a significant defect.
This function is more simple and easier to handle than that Fock's one. Probably the first  attempt to use these coordinates directly in variational
 approach was done in \cite{RefMorre} , but authors did not use explicitly correlated functions.\\
As a first test of quality of the basis set all calculations were done with the same exponential scale 
factor $\zeta$ in $\exp{(-\zeta R)}$. 
An important feature of this approach is inclusion of negative powers of $R$ with combination of positive powers of $r_{12}$.
Incorporation of these terms is -- 
similar to the approach of Kinoshita $[3]$ -- 
necessary for the wave function to be a formal solution of 
Schr\"odinger equation. 
As a first step the variational method has been used. We want to show that this (or similar)  proposed basis sets open new
possibilities in finding of the proper wave function that satisfies all cusp conditions.  
We believe that this knowledge will increase a chance to construct general few electron (with number of electrons more than two) atomic (or molecular) wave functions, because of 
a pairwise character of electron-electron and nuclear-electron interactions. 

\section{Hamiltonian transformation and basis set construction}

The Hamiltonian for Helium atom in S basic state 
in coordinates $r_{1}$, $r_{2}$ and $r_{12}$ reads
%
\begin{eqnarray}
&H=&
- \frac{1}{2}\frac{\partial^{2}}{\partial r_{1}^{2}}-\frac{1}{r_{1}}\frac{\partial}{\partial r_{1}}
 -\frac{1}{2}\frac{\partial^{2}}{\partial r_{2}^{2}}-\frac{1}{r_{2}}\frac{\partial}{\partial r_{2}}
\nonumber
\\
&&- \frac{\partial^{2}}{\partial r_{12}^{2}}-\frac{2}{r_{12}}\frac{\partial}{\partial r_{12}}
-\frac{1}{2}\frac{r^{2}_{1}+r^{2}_{12}-r^{2}_{2}}{r_{1}r_{12}}\frac{\partial^{2}}{\partial r_{1}\partial r_{12}}
-\frac{1}{2}\frac{r^{2}_{2}+r^{2}_{12}-r^{2}_{1}}{r_{2}r_{12}}\frac{\partial^{2}}{\partial r_{2}\partial r_{12}}
\nonumber
\\
&&- \frac{Z}{r_{1}}-\frac{Z}{r_{2}}+\frac{1}{r_{12}}
\label{Hamiltonian}
\end{eqnarray}
As it was already mentioned in the introduction, a function created 
from only the powers of $r_{1}$, $r_{2}$ and $r_{12}$ 
is not an exact eigenfunction of the Hamiltonian (\ref{Hamiltonian}) 
due to the cross terms 
$
\frac{r^{2}_{2}}{r_{1}r_{12}}\frac{\partial^{2}}{\partial r_{1}\partial r_{12}}
$ 
and
$
\frac{r^{2}_{1}}{r_{2}r_{12}}\frac{\partial^{2}}{\partial r_{2}\partial r_{12}}
$.
 Let us make the following transformation of the coordinate system:
\begin{eqnarray}
R=\sqrt{r_{1}^{2}+r_{2}^{2}}
\\
t=\frac{r_{2}^{2}-r_{1}^{2}}{r_{2}^{2}+r_{1}^{2}}
\end{eqnarray}
so the $r_{1}$ and $r_{2}$ coordinates can be expressed as
\begin{eqnarray}
r_{1}=\frac{R \sqrt{1-t}}{\sqrt{2}}\nonumber\\
\\
r_{2}=\frac{R \sqrt{1+t}}{\sqrt{2}}
\end{eqnarray}

After somehow lengthy manipulations we obtain the following transformed Hamiltonian 
%
\begin{eqnarray}
&H=&
 - \frac{1}{2}\frac{\partial ^{2}}{\partial R^{2}} - \frac{5}{2R}\frac{\partial}{\partial R}
- \frac{2(1-t^{2})}{R^{2}}\frac{\partial^{2}}{\partial t^{2}}
+ \frac{6t}{R^{2}}\frac{\partial}{\partial t} 
\nonumber
\\
&&
 - \frac{r_{12}}{R}\frac{\partial^{2}}{\partial R\partial r_{12}}
- \frac{2t}{r_{12}}\frac{\partial^{2}}{\partial t\partial r_{12}}
+ \frac{2t.r_{12}}{R^{2}}\frac{\partial^{2}}{\partial t\partial r_{12}}
\nonumber
\\
&&
 - \frac{\partial^{2}}{\partial r^{2}_{12}} - \frac{2}{r_{12}}\frac{\partial}{\partial r_{12}} -\frac{Z\sqrt{2}}{R\sqrt{1-t}} -\frac{Z\sqrt{2}}{R\sqrt{1+t}}+\frac{1}{r_{12}}
\label{H_transf}
\end{eqnarray}
We see that the cross terms that caused problems in (\ref{Hamiltonian}) 
are now transformed to somewhat more convenient expressions, 
namely, the differential terms now always contain at most 
a single singularity factor [compared with double singularity factors 
$(r_1 r_{12})^{-1}$, $(r_2 r_{12})^{-1}$  in (\ref{Hamiltonian})] 
with all the unconvenient double singularity factors occuring now 
only in potential terms.
It seems rather natural to include some combinations of the powers of 
$R$, $r_{12}$ and $t$ 
into the wave function.
Moreover, we must carefully treat the singularities in potentials 
and the antisymmetry of the wave function.  
If we could expand the potential term 
\begin{equation}
V_{nuc} =
\frac{Z \sqrt{2}}{R}(\frac{1}{\sqrt{1-t}}+\frac{1}{\sqrt{1+t}})
\label{V_nuc_1}
\end{equation}
in a Taylor series around $t=0$ in the whole range of $t$, 
the solution in the form of linear combinations of 
$R^i r_{12}^j t^{2k} exp(-\zeta R)$ 
would indeed be sufficient. 
Our analysis showed (see Appendix) that in this case we can eliminate 
all the singularities when we also include 
-- similar to Kinoshita $[3]$ -- 
combinations with negative exponents $i$ for $R^i$. 
Restricting the Taylor expansion to even powers of $t$ guarantees the required 
symmetry with respect to the particle interchange, 
moreover, odd powers of $t$ mutually cancel 
in the Taylor expansion of (\ref{V_nuc_1}) anyhow.
However, this Taylor expansion evidently diverges for $t=\pm 1$.
The modified expansion
\begin{equation}
V_{nuc} =
 -\frac{2\sqrt{2}Z}{R\sqrt{1-t^{2}}}(1-\sum_{k=1}^{\infty}\frac{(4k-3)!!}{4k!!}t^{2k})
\label{V_nuc_2}
\end{equation}
already correctly reproduces the singular behaviour for $t=\pm 1$, though, 
the convergence of the series is slower in the vicinity of the singularities. 
On the other hand, the alternative expansion 
\begin{equation}
V_{nuc}=-\frac{2Z}{R\sqrt{1-t^{2}}}\Big(1+\frac{1}{2}\sqrt{1-t^{2}}+\sum_{k=2}^{\infty}\frac{(-1)^{k+1}(2k-3)!!}{2k!!}\sqrt{1-t^{2}}^{k}\Big)
\label{V_nuc_3}
\end{equation}
works well anywhere except the vicinity of $t=0$.
Based on these potential expansion considerations 
we suggested the following form for our wave function (para case)
\begin{eqnarray}
\Psi_{1}= \sum_{i,j}^{\infty}\sum_{k=0}^{\infty}\sum_{l=0}^{1}
C_{ijkl}R^{i}r^{j}_{12}t^{2k}\sqrt{1-t^{2}}^{l} e^{-\zeta R}
\frac{\alpha(1)\beta(2)-\alpha(2)\beta(1)}{\sqrt{2}}.
\label{Psi1}
\end{eqnarray}
This form of a function has already a chance to eliminate the potential term 
$1/\sqrt{1-t^2}$. 
Summations over $l$ in (\ref{Psi1})
are restricted to $0$ and $1$ in order to avoid redundancies 
caused by combinations of powers of $t^{2k}$.

By linear combinations of the functions of the type (\ref{Psi1}) 
a formally simpler form can be obtained
\begin{eqnarray}
\Psi_{2}= \sum_{i,j}^{\infty}\sum_{k=0}^{\infty}C_{ijk}R^{i}r^{j}_{12}\sqrt{1-t^{2}}^{k} e^{-\zeta R}\frac{\alpha(1)\beta(2)-\alpha(2)\beta(1)}{\sqrt{2}}.
\label{Psi2}
\end{eqnarray}
As (\ref{Psi1}) can be transformed into (\ref{Psi2}) 
it is not surprising that also for a finite truncation 
of infinite sums in (\ref{Psi1}), (\ref{Psi2}) 
these two functions give for a comparable basis sets 
almost the same results. 
In our test calculations the wave functions of the type (\ref{Psi1}) 
gave slightly better results, so we decided to work further with $\Psi_1$ 
rather than $\Psi_2$ 
(we will refer to this type of function as Basis A in the text). 

For even better cancellation of the nuclear potential we suggested 
a wave function of a little more complex form,
\begin{eqnarray}
\Psi_{3}=\Big( \sum_{i}^{\infty} \sum_{j=0}^{\infty} R^{i} r^{j}_{12} e^{-\zeta R}\big(\sum_{k=0}^{\infty}A_{i,j,k}t^{2k}(\sqrt{1-t}+\sqrt{1+t})+
\nonumber\\
\sum_{k=1}^{\infty}B_{i,j,k}t^{2k-1}(\sqrt{1-t}-\sqrt{1+t})\big)+
\nonumber\\
\sum_{i}^{\infty}\sum_{j=0}^{\infty}\sum_{k=0}^{\infty}
R^{i}r^{j}_{12} e^{-\zeta R}\sum_{l=0}^{1}C_{ijkl}t^{2k}\sqrt{1-t^{2}}^{l}\Big) 
\frac{\alpha(1)\beta(2)-\alpha(2)\beta(1)}{\sqrt{2}}.
\label{Psi3}
\end{eqnarray}
Expansion in this basis set (denoted as Basis B) has the advantage that 
for properly chosen coefficients $A_{ijk},B_{ijk},C_{ijkl}$ 
acting with (\ref{H_transf}) on (\ref{Psi3}) results in the same type of 
functions on both sides of the Schr\"odinger equation.
Up till now we found nothing that would contradict the statement 
that (\ref{Psi3}) is a formal solution of the Schr\"odinger equation.

In addition, we considered also a modification of (\ref{Psi3}) 
with more complex exponential factor  
$\exp\{-\zeta R(\sqrt{1-t}+\sqrt{1+t})/\sqrt{2}\}$
instead of the simple $\exp(-\zeta R)$.  
This function we denoted as Basis C.

Notice that each of the suggested forms of wave function 
(\ref{Psi1}) - (\ref{Psi3})  
is also automatically antisymmetric with respect to particle interchange.

Substitution of $\Psi$ [either in the form 
(\ref{Psi1}), (\ref{Psi2}) or (\ref{Psi3})]  
into the Schr\"o\-din\-ger equation 
results in a set of coupled equations for the coefficients $C$ 
(or, $A$, $B$, $C$ in the case of $\Psi_3$). 
Direct solution of these sets of equations is (similarly as was done in excelent work of Pekeris \cite{RefPek})
, however,  very difficult, so as a first step 
the standard variational approach was used.

\section{Variational solution of Schr\"odinger equation}
Here we solved the standard matrix equation
\begin{equation}
\mathbf{H} \mathbf{c} = \epsilon \mathbf{S}\mathbf{c}
\end{equation}
The elements of the matrix $\mathbf{H}$ are $\langle \Phi_{i',j',k',l'} \mid H \mid \Phi_{i,j,k,l} \rangle$
and that of the overlap matrix $\mathbf{S}$ are $\langle \Phi_{i',j',k',l'} \mid \Phi_{i,j,k,l}\rangle$.
Finally the vectors $\mathbf{c}$ consist of the coefficients $C_{i,j,k,l}$ in case of basis A  
and
of $A_{i,j,k}$, $B_{i,j,k}$ and $C_{i,j,k,l}$ in the case of bases B and C. 

\subsection{Calculation of integrals}
When we use coordinates $r_{1}$, $r_{2}$ and $r_{12}$ volume of the integration is $8\pi^{2} r_{1}r_{2}r_{12}\\
.dr_{1} dr_{2}dr_{12}$. 
After a series of lengthy manipulations we succeeded to show that acting with Hamilton operator on $\Phi_{i,j,k,l}$ results in some linear combination of the type $\Phi_{i',j',k',l'}$ again. Consenquently, if we use function $\Psi_{1}$ or $\Psi_{3}$ the only type of integral which we will encounter when solving (10) will be of the type 
\begin{eqnarray}
8\pi^{2} \int \Phi_{i,j,k,l}r_{1}r_{2}r_{12}dr_{1} dr_{2}dr_{12}=\nonumber\\
=8\pi^{2}\int \Phi_{i,j,k,l}(R,t,r_{12})J_{\phi(R,t)}R^{2}\frac{\sqrt{1-t^2}}{2}dR dt dr_{12}\nonumber\\
\end{eqnarray}
The Jacobian of the transformation $r_{1},r_{2} \to R,t$ is $J_{\phi(R,t)}=\frac{R}{2\sqrt{1-t^2}}$.
Evidently, this transformation is regular except of the point $r_{1}=0,r_{2}=0$. Integration limits of the $r_{12}$ 
coordinate do not change, thus the region of the integration must be divided into two parts, 
namely into $t\in \langle -1,0 \rangle $, (the $r_{1}>r_{2}$ case) and
the region $t\in \langle 0,1 \rangle$, (the $r_{2}>r_{1}$ case). 
The region of integration of 
coordinate $R$ is independent  so when we considered the integrals of kind
$\langle \Psi \mid \Psi \rangle$ or $\langle \Psi \mid H \mid \Psi \rangle$
 (here $\Psi$ could be $\Psi_{1}$, $\Psi_{2}$ or $\Psi_{3}$) we  
have  these two types of integrals:
\begin{displaymath}
\mathcal{I}_{1}=2\pi^{2}\int\limits_{0}^{\infty}R^{3}dR\int\limits_{0}^{1}dt\int\limits_{\big | R\sqrt{\frac{1-t}{2}}-R\sqrt{\frac{1+t}{2}}\big |}^{R\sqrt{\frac{1-t}{2}}+R\sqrt{\frac{1+t}{2}}}r_{12}f(R,t,r_{12})dr_{12} \qquad r_{2}>r_{1}
\end{displaymath}
\begin{displaymath}
\mathcal{I}_{2}=2\pi^{2}\int\limits_{0}^{\infty}R^{3}dR\int\limits_{-1}^{0}dt\int\limits_{\big | R\sqrt{\frac{1-t}{2}}-R\sqrt{\frac{1+t}{2}} \big |}^{R\sqrt{\frac{1-t}{2}}+R\sqrt{\frac{1+t}{2}}}r_{12}f(R,t,r_{12})dr_{12} \qquad r_{1}>r_{2}
\end{displaymath}
 where 
\begin{displaymath}
f(R,t,r_{12})=R^{I}r^{J}_{12}t^{K}\sqrt{1-t^{2}}^L\sqrt{1-t}^M\sqrt{1+t}^N\exp{2(-\zeta R)}
\end{displaymath}
and in case of basis $C$ $f(R,t,r_{12})$ is the same function 
the only difference is that it has the exponential function 
$\exp\{-\zeta R(\sqrt{1-t}+\sqrt{1+t})/\sqrt{2}\}$. 
Index K is always positive and I,J,L,M and N integers could be also negative. 
We integrate through coordinate $r_{12}$ first:
\begin{eqnarray}
\int\limits_{\big | R\sqrt{\frac{1-t}{2}}-R\sqrt{\frac{1+t}{2}} \big |}^{R\sqrt{\frac{1-t}{2}}+R\sqrt{\frac{1+t}{2}}}r_{12}^{J+1}dr_{12} = 
\left\{ \begin{array}{lll}
\frac{2}{J+2}(\frac{R}{\sqrt{2}})^{J+2}\sum_{i=1}^{J'}{J+2 \choose i}(1-t)^{\frac{J+2-i}{2}}(1+t)^{\frac{i}{2}} & \textrm{if $r_{2}>r_{1}$}
\nonumber\\
\nonumber\\
\frac{2}{J+2}(\frac{R}{\sqrt{2}})^{J+2}\sum_{i=1}^{J'}{J+2 \choose i}(1+t)^{\frac{J+2-i}{2}}(1-t)^{\frac{i}{2}} & \textrm{if $r_{1}>r_{2}$}
\nonumber\\
\end{array} \right.
\end{eqnarray}
where $J'$ is largest odd number from $J+2$.
We can change the order of integration and integrate through coordinate $R$ so in case A or B we finally tackle the integrals of type 
\begin{eqnarray}
\mathcal{I}=\int\limits_{0}^{1}t^{K}\sqrt{1-t^{2}}^L\sqrt{1-t}^{M}\sqrt{1+t}^{N}dt,
\label{Int1}
\end{eqnarray}
and when we are using basis sets of type C, we have integrals of type  
\begin{eqnarray}
\mathcal{I}=\int\limits_{0}^{1}\frac{(t^{K}\sqrt{1-t^{2}}^L\sqrt{1-t}^{M}\sqrt{1+t}^{N}}{(\sqrt{1-t}+\sqrt{1+t})^{O}}dt.
\label{Int2}
\end{eqnarray}
The analytical solutions of integrals of the type (\ref{Int1}) and (\ref{Int2}) exists. 

A software package for evaluation of these integrals in Fortran90 language 
was written by the author. 
Calculations were done in real*16 (128-bit) arithmetics 
as the usual real*8 (64-bit) arithmetics 
was found to be unsufficient here.
Our experience showed that for more precise calculations 
even the use of higher precision (real*24 or real*32) would be advisable.
Almost all calculations were done on a PC with 2.8GHz processor 
with 4GB RAM memory under Linux operating system. 
The calculation of the integrals was made more effective by pre-calculating 
of the integrals (\ref{Int1}) or (\ref{Int2}),
so for example for the A-type basis set with $I_{max}=5$ 
calculation of all integrals of type (14) 
on  this PC is a question of few seconds. 
\\
The most time consuming part of the calculation remained 
the diagonalization of the overlap and transformed Hamiltonian matrices.

\subsection{Details and results}

The wave functions of the type 
$\Psi_{1}$, $\Psi_{2}$ and $\Psi_{3}$ 
(\ref{Psi1}-\ref{Psi3})  
were tested in our calculations.
In practice, the infinite summations over $i,j,k$  
in (\ref{Psi1}-\ref{Psi3})  
have to be obviously truncated to finite values 
$I_{m},J_{m},K_{m}$, respectively. 
Unfortunately, for larger powers of $k$ 
our basis sets are close to linear dependency.  
We used standard canonical transformation of 
the basis set with elimination of small eigenvalues  below a defined threshold 
to overcome this problem.
In all cases, the choice of the threshold was governed by the 
criterion to be the minimal threshold where the method still converges. 
Exact nuclear potential $-Z/r_{i}$ $(i=1,2)$  was used 
in our variational calculations. 
For each $\Psi_i$ $(i=1,2,3)$ such exponents $\zeta$ were chosen that 
$E=E(\zeta)=\langle \Psi_i(\zeta) \vert H \vert \Psi_i(\zeta) \rangle$ 
is minimal with respect to $\zeta$. 
To find the minimum of $E(\zeta)$ the golden section technique was used.  
Results of our calculations were compared with 
reference values for nonrelativistic energy of the ground state of He atom 
and H$^{-}$ ion 
according to \cite{RefSim}.

We used Basis A functions 
[$\Psi_1$ and $\Psi_2$, Eqs. (\ref{Psi1}-\ref{Psi2})] first. 
It turned out that there are almost no differences between using basis sets 
$\Psi_1$ and $\Psi_2$ of equivalent sizes. 
This is not surprising as $\Psi_1$ can be in principle generated from $\Psi_2$ 
and vice versa when untruncated infinite summations are considered. 
Our numerical experience showed that the truncated summations 
with $\Psi_1$ produced slightly better results than with $\Psi_2$. 
Thus, all our results presented with Basis A refer to 
using $\Psi_1$. These are summarised in Tab.1. 
Each combination of $I_{m},J_{m},K_{m}$ determines 
the total number of basis functions, $N_0$. 
By $N$ the number of basis set functions 
after elimination of quasi-linear dependencies is denoted. 
This represents then the actual basis set size used in the calculation. 
However, as $N$ is unknown before we start the calculation, 
$N_0$ was used to label our basis sets 
(preceded by basis type label A,B, or C). 
The value of optimal exponent $\zeta$ for each basis set is presented as well. 

As emphasised in the Appendix, negative powers of $R$ 
[in combination with positive powers of $r_{12}$ according to (\ref{App_r12Rlim}) 
i.e., $r_{12}^{j}/R^{i'}$ up to $i'=j$]  
should be included to eliminate the singularities of the type $R^{-2}$. 
Unless explicitly specified otherwise, 
all the basis sets used in our calculations were constructed in this way 
(both for A and B/C case). 
To see what happens when we do not include negative powers of $R$ in 
the wave function a few illustrative calculations with only positive powers 
of $R$ are included. 
For basis sets presented in Tab.1 this is the case of the set A438a. 
Almost no effect on the ground state energy can be seen for this basis set. 
The effect will be more visible for the basis sets B and C, 
especially for the cusp parameters.

To test the stability of the method the hydrogen anion energy 
was considered as well.
Evidently, the H$^{-}$ anion is a more sensitive system. 
Here we found to be necessary to use 
the strict threshold of $10^{-7}$ for elimination of linear dependencies. 
(For He a threshold of $10^{-12}$ was found to be sufficient 
for A-type basis sets and even $10^{-16}$ for basis sets B and C.)
The difference in minimal threshold for He and H$^{-}$ anion explains why our energies for He 
are closer to the reference value than those for the H$^{-}$ anion. 

From Tab.1 we see that 
even with the simple basis set functions of the type (\ref{Psi1}) 
the He and H$^{-}$ ground state energy with the accuracy of $10^{-8}$ 
can be achieved with a moderate basis set size. 

Evidently, $\Psi_1$ and $\Psi_2$ can be the exact solutions 
only for the approximate potential (\ref{V_nuc_2}) or (\ref{V_nuc_3}). 
Therefore, basis sets of the type B or C 
which satisfy the exact Coulomb potential 
are expected to produce better results.
In Tab.2 results on a hierarchy of these basis sets is presented. 
When comparing the energies from Tab.1 and Tab.2 we see that the basis sets 
of the type B and C produce indeed better energies than the A set case. 
The best result was obtained with basis set C1068. 
For comparison we present also results with two modified 
non-standard way generated sets.
The set C944a contains only positive powers of $i$ with $I_{m}=7$ 
($I_{m}=6$ in all other basis sets with $N_{0}>900$). 
The basis set C950b contained negative powers of $i$ with $I_{min}=-J_{m}+2$ which corresponds to the constraint that was put on the function in 
Appendix (\ref{App_PhiJ}).
\\
If we compare the energies corresponding to basis sets C944a and C950b 
(which are of almost the same size) 
the energy of C950b is  by one order better than the basis set C944a
which confirms the importance of inclusion of negative powers of $R$. 
As the practice shows
incorporation of negative powers of $R$ in combination $r_{12}^{j}/R^{i}$ 
is at least as important as the inclusion of higher positive powers of $R$.  

To assess the behaviour of our wave functions in the vicinity of 
the electron-nuclear and electron-electron coalescence point 
the cusp parameters $\nu_1$ and $\nu_{12}$ were evaluated,
\begin{eqnarray}
\nu_{1} &=& 
\frac{\langle \delta(\mathbf{r}_{1})\frac{\partial}{\partial r_{1}} \rangle}{\langle \delta(\mathbf{r}_{1}) \rangle}
\label{cusp1}
\\
\nu_{12} &=& 
\frac{\langle \delta(\mathbf{r}_{12})\frac{\partial}{\partial r_{12}} \rangle}{\langle \delta(\mathbf{r}_{12}) \rangle}
\label{cusp2}
\end{eqnarray}
where
$\langle{\bf O}({\bf r}_1 ,{\bf r}_2 )\rangle = \int d{\bf r}_1 d{\bf r}_2 \,
\Psi({\bf r}_1 ,{\bf r}_2 )^{*} \,
{\bf O}({\bf r}_1 ,{\bf r}_2 ) \, 
\Psi({\bf r}_1 ,{\bf r}_2 ) \, $ 
and $\delta(\mathbf{r})$ is Dirac delta function.
Result are summarised in Tab.3. 
As expected, the best agreement was obtained for the C-type  
basis set C1068.
The agreement for A-type basis sets was for about two orders worse. 
Though it is not always justified to compare the A,B and C basis sets directly, 
these results indicate that better wave functions can be obtained with B and C sets.

\section{Discussion}

The basis sets could be furthet optimised by omitting the contributions with 
negligible effect on the result. Therefore, it would probably be possible 
to find different basis sets with the same total number of functions 
that could result in better energy.
We did not follow this way. 
Let us notice that almost for each basis for small values of $\zeta$ 
a "grey area" exists where the method does not converge 
due to linear dependency problems. 
This is more visible for H$^{-}$ anion. 
\\
It is clear, that functions of type $t^{2k}$ and  $t^{2k+2}$ or $t^{2k+1}$ and  $t^{2k+3}$, 
for large $k$ tend to be similar within the interval $\langle -1,1 \rangle$. 
One can easily check that the set $t^{2k}, k=0,1,\dots,K_m$ 
is already for $K_m \approx 10$
almost linearly dependent on the interval $\langle -1,1 \rangle$. 
Evidently, this limits the basis set size in practice and, consequently, also the precision of our energy (or any other quantity of interest) calculation. 
From this point of view, basis sets B and C are again superior to A as using the same $K_m$ higher precision can be obtained with B or C than with A type basis set (compare Tab.1. and Tab.2.). 
A remedy how to increase the precision for B or C sets 
keeping $K_m$ the same would be to include more exponents 
$\zeta_{n}$: $\exp{-\zeta_{n}R}$ or  $\exp{-\zeta_{n}.(r_{1}+r_{2})}$. 
Another possibility would be to use Fock logarithmic terms $R^{i} \log{R}$
 to better describe the triple coalescent point of the wave function 
(i.e., $r_{1}=r_{2}=0$). The large advantage of basis sets of type 
B is the possibility of exact calculation of integrals of the type
\begin{equation}
I_{p}(i,\zeta)= \int_{0}^{\infty}dR R^{i}(\log{R})^{p}\exp{-\zeta.R}\\
\end{equation}
contrary to the methods that use Hylleraas coordinates.
In this case, for the basis of the C type 
an analytical solution for the integrals over $t$ probably does not exist 
and the integrals have to be evaluated numerically. 
An alternative would be a combination of the method with more exponents 
with Fock logarithmic terms.
We expect that each of the aformentioned procedures leads to 
improvement of basis set linear dependency problems and 
a higher precision of calculations as a result.


\begin{table*}
\caption
{
Ground state energies of He and H$^{-}$ (at.units.) 
using wave functions of the type (\ref{Psi1}). 
Number of basis functions after elimination of linear dependencies $N$, 
infinite summation truncations $I_{m}, J_{m}, K_{m}$ 
and optimised exponents $\zeta$ for each basis set are also shown. 
}
\begin{tabular}{|l|c|c|c|c|c|c|}
\hline
\multicolumn{7}{|l|}{\quad He}\\
\hline
 Basis set & \ \ $N$ \ \ & $I_{m}$ & $J_{m}$ & $K_{m}$ & $\zeta$ & Energy \\
\hline
 A160     & 100 & 3 & 3 & 5 & 2.429563 &   $ -2.90372198 $    \\
 A438     & 231 & 5 & 5 & 5 & 2.460426 &   $ -2.90372427 $    \\
 A438a    & 236 & 6 & 5 & 5 & 2.680000 &   $ -2.90372427 $    \\
 A860     & 456 & 7 & 7 & 5 & 2.249919 &   $ -2.90372433 $    \\
\hline
\multicolumn{6}{|l|}{Exact$^a$ } &
\multicolumn{1}{|c|}{$ -2.90372438 $} \\
\hline
\multicolumn{7}{|l|}{\quad H$^-$}\\
\hline
  A860 & 530 & 7 & 7 & 5 & 0.851281 & $ -0.52775094 $  \\
\hline
\multicolumn{6}{|l|}{Exact$^a$ } &
\multicolumn{1}{|c|}{$ -0.52775102 $} \\
\hline
\multicolumn{1}{l}{ $^a${\small{Ref.\cite{RefSim}}}}
\end{tabular}
\end{table*}


\begin{table*}
\caption
{
Ground state energy (at.units) of He atom and H$^{-}$ ion
using (\ref{Psi3}).     
}
\begin{tabular}{|l|c|c|c|}
\hline
\multicolumn{4}{|l|}{\quad He}\\
\hline
 Basis set & \ \ $N$ \ \ & $\zeta$ & Energy \\
\hline
 B288  & 137   & 2.300232 & $ -2.9037240795  $   \\
 C288  & 129   & 2.063932 & $ -2.9037242527  $   \\
 B595  & 222   & 2.197048 & $ -2.9037243113  $   \\
 C595  & 193   & 2.182277 & $ -2.9037243717  $   \\
 B1068 & 404   & 2.750647 & $ -2.9037243755  $   \\
 C1068 & 390   & 2.342957 & $ -2.9037243767  $   \\
\hline
 C944a  & 351  & 2.484067 & $ -2.9037243750  $   \\
 C950b  & 366  & 2.339823 & $ -2.9037243766  $   \\
\hline
\multicolumn{3}{|l|}{Exact$^a$ } &
\multicolumn{1}{|l|}{$-2.9037243770$} \\
\hline
\multicolumn{4}{|l|}{\quad H$^-$}\\
\hline
 C1068 & 358   & 0.738130 & $ -0.5277509907 $    \\
\hline
\multicolumn{3}{|l|}{Exact$^a$ } &
\multicolumn{1}{|l|}{$-0.5277510165$} \\
\hline
\multicolumn{1}{l}{ $^a${\small{Ref.\cite{RefSim}}}}
\end{tabular}
\end{table*}


\begin{table*}
\caption
{
Cusp parameters [see Eqs.(\ref{cusp1}-\ref{cusp2})]            
}
\begin{tabular}{|l|c|c|l|l|}
\hline
 Basis set &  $\nu_{1}$ & $\nu_{12}$ & 
 \ $\langle \delta(\mathbf{r}_{1})\rangle$ & 
 \ $\langle \delta(\mathbf{r}_{12})\rangle$ \\
\hline
A438a       & -1.974804 & 0.491942 & 1.809316 & 0.106425 \\
A438        & -1.976622 & 0.492143 & 1.809494 & 0.106435 \\
A860        & -1.985193 & 0.494084 & 1.809791 & 0.106400 \\
B1068       & -1.999224 & 0.498531 & 1.810405 & 0.106349 \\
C1086       & -2.000097 & 0.499408 & 1.810432 & 0.106348 \\
\hline
C944a       & -2.000034 & 0.496380 & 1.810429 & 0.106372 \\
C950b       & -2.000014 & 0.498969 & 1.810429 & 0.106351 \\
\hline
Ref.values & $-2.^{a}$  & $0.5^{a}$  & $1.810429^{b}$ & $0.106345^{b}$ \\
\hline
\multicolumn{1}{l}{ $^a${\small{Exact, see \cite{RefKato}}}}\\
\multicolumn{1}{l}{ $^b${\small{Ref.\cite{RefBDrake}}}}
\end{tabular}
\end{table*}

\section{Conclusions}
In this work a new ansatz for the Helium-like ion wave function  was  proposed. 
To our knowledge it is at present the best energy for He atom obtained 
with the use of hyperradial coordinates.
Our preliminary results show that the wave functions of the type B and C 
are close to the exact solution of the Schr\"odinger equation.
The remarkable property of the functions B and C is that for properly chosen 
expansion coefficients functions of the same type occur on both sides of the 
Schr\"odinger equation. 
In \cite{RefDru} it was shown that a simple exponential 
$\exp{-2\zeta R}$ describes the behaviour of the wave functions fairly well 
in limit cases $r_{1}=0$, $r_{2}=0$  or $r_{12}=0$. 
We showed that the functions A or B 
(containing just the aforementioned exponential function)
can offer relatively precise ground state energies.
However, the calculations of cusp parameters indicate 
that the use of exponentials $\exp{-2\zeta (r_{1}+r_{2})}$ 
should be prefered. 
The important point turnes out to be the inclusion of negative powers of 
$R$ in combination with positive powers of $r_{12}$
similar to Kinoshita \cite{RefKin} where negative powers of 
$s=r_{1}+r_{2}$ were included. 
Incorporation of negative powers of $R$ is at least as important as the use of 
larger number of positive powers of $R$. 
For our method the inclusion 
of the full correction to finite mass 
of the nuclea (see e.g.\cite{RefBBethe} p.263) is straightforward, 
the work is in progress. 
Our method in its present state could not compete with ICI method 
of Nakatsuji \cite{RefNak1,RefNak2,RefNak3}, 
but our preliminary results urge us to clousure that 
the wave function of type (\ref{Psi3}) 
is close to exact one and one of proposed ways of improvement 
of the wave function  could produce more exact functions. 
The knowledge of proper functional behavior of three-body wave 
function could be interesting also for the ICI method.

Finally, there was done some work on generalization of this ansatz to systems 
with more than two electrons, we believe that we are on the right way to
overcome the problems with multielectron integrals that generally depend 
on $r_{ij}$.

\section{Appendix}

\setcounter{equation}{0}
\renewcommand{\theequation}{A-\arabic{equation}}
In this appendix we will show that inclusion 
of terms $R^{-i}$ 
(up to $i_{max}=j$) 
is a necessary condition for a function
\begin{equation}
\Psi = \sum_{j=0}^{\infty} \Phi_{j}(R,t)r_{12}^{j}
\label{App_Psi}
\end{equation}
to be (a spinless part of) a correct solution of our problem.
We will prove this for a case where we can assume 
that $V_{nuc}$ ({\bf{8}}) 
can be expanded in a Taylor series around $t=0$,
\begin{equation}
V_{nuc} = -\frac{2 Z \sqrt{2}}{R} 
\big ( 1+\sum_{k=1}^{\infty}\frac{(4k-1)!!}{(4k)!!} t^{2k} \big ).
\label{App_Vnuc}
\end{equation}

Let us act with the Hamilton operator (\ref{H_transf}) on a function (\ref{App_Psi}). 
We obtain a series of coupled equations  
\begin{eqnarray}
\Big((j+2)(j+3)+2(j+2)t\frac{\partial}{\partial t}\Big )\Phi_{j+2}=\Big (H_{j}-E\Big )\Phi_{j}+\Phi_{j+1}
\label{App_Phi}
\end{eqnarray}
where the operator $H_{j}$ is 
\begin{eqnarray}
H_{j}= - \frac{1}{2}\frac{\partial ^{2}}{\partial R^{2}} - \frac{5}{2R}\frac{\partial}{\partial R}
- \frac{2(1-t^{2})}{R^{2}}\frac{\partial^{2}}{\partial t^{2}}
+ \frac{6t}{R^{2}}\frac{\partial}{\partial t} \nonumber\\
- \frac{j}{R}\frac{\partial}{\partial R}
+ \frac{2jt}{R^{2}}\frac{\partial}{\partial t}
-\frac{Z\sqrt{2}}{R\sqrt{1-t}} -\frac{Z\sqrt{2}}{R\sqrt{1+t}}
.
\label{App_Hj}
\end{eqnarray}
Setting $\Phi_j = 0$ for all $j<0$ the first two equations of the set (\ref{App_Phi}) read
\begin{eqnarray}
(2+2 t\frac{\partial}{\partial t})\Phi_{1} &=& \Phi_{0} 
\label{App_Phi0}
\\
(6+4 t\frac{\partial}{\partial t})\Phi_{2} &=& (H_{0}-E)\Phi_{0}+\Phi_{1}
\label{App_Phi1}
\end{eqnarray}
Substituting for $\Phi_0$ from (\ref{App_Phi0}) into (\ref{App_Phi1}) we get
\begin{eqnarray}
(1+\frac{2}{3} t\frac{\partial}{\partial t})\Phi_{2}=
\frac{1}{3}(H_{0}-E)(1+t \frac{\partial}{\partial t})\Phi_{1}
+\frac{1}{6}\Phi_{1}
\label{App_Phi2}
\end{eqnarray}
from which we can get a formal solution
\begin{eqnarray}
\Phi_{2}=\frac{1}{t^{3/2}}\int_{0}^{t}  t'^{1/2}
\big [\frac{1}{2}(H_{0}-E)(1+t'\frac{\partial}{\partial t'})+\frac{1}{4}\big ] 
\Phi_{1} dt'. 
\label{App_Phi2int}
\end{eqnarray}
Generally, a recursive formula for a solution of (\ref{App_Phi}) can be found,
\begin{eqnarray}
\Phi_{j+2}=\frac{1}{t^{(j+3)/2}}\int_{0}^{t} t'^{(j+1)/2}\frac{1}{2(j+2)}
\big [(H_{j}-E)\Phi_{j}+\Phi_{j+1}\big ] dt', 
\qquad \quad j=0,1,\dots
\label{App_PhiRec}
\end{eqnarray}
indicating that, using (\ref{App_Phi2int}) and (\ref{App_PhiRec}) 
all $\Phi_j$'s can be, in principle, obtained. 

A natural requirement imposed on $\Phi_j$'s is that $\Phi_j r_{12}^{j}$ [see(1)] are expected to be all finite and decaying to zero with increasing $j$.
\\
As we here consider the situation where the expansion (\ref{App_Vnuc}) can be used we can suggest $\Phi_j$'s to be of the form
\begin{eqnarray}
\Phi_{j}=\sum_{i,k} C_{i,j,k} R^{i} t^{2k} e^{-\zeta R}. 
\label{App_PhiJ}
\end{eqnarray}
Acting with the Hamilton operator on (\ref{App_PhiJ}) can, however, 
produce singularities of the type $R^{-1}$, $R^{-2}$. 
Evidently, the suggested form (\ref{App_PhiJ}) 
must be able to eliminate all these singularities.
Let us inspect this more closely.

Substituting (\ref{App_PhiJ}) into (\ref{App_Phi2int}) yields
\begin{eqnarray}
\Phi_{2}=\sum_{i,k} C_{i,1,k}(1+2k)\Big ( \frac{1}{3+4k}\big (-(\frac{1}{2}\zeta^{2}+E)R^{i}+\nonumber\\
+\zeta(i+\frac{5}{2})R^{i-1}-\frac{i(i+4)}{2}R^{i-2}
+2(2k)(2k-1)R^{i-2}+12k R^{i-2}-\nonumber\\
-2\sqrt{2}Z R^{i-1}\big )t^{2k}-\frac{1}{3+4k-4} 2(2k)(2k-1)R^{i-2}t^{2k-2}-\nonumber\\
-2\sqrt{2}Z R^{i-1}\sum^{\infty}_{p=1}\frac{1}{3+4(k+p)}\frac{(4p-1)!!}{(4p)!!}t^{2(k+p)}\Big )+\nonumber\\
+\sum_{i,k} C_{i,1,k}\frac{1}{2}\Big (\frac{1}{3+4k}R^{i}t^{2k}\Big )
\label{App_Phi2_Int}
\end{eqnarray}

For elimination of all singularities of the type $R^{-2}$ from $\Phi_{2}$  
it is evidently sufficient to put 
\begin{eqnarray}
C_{0,1,k}=0,
\qquad \qquad \qquad k=1,2,\dots
\label{App_R2elim}
\end{eqnarray}
To eliminate the $R^{-1}$ singularities from $\Phi_2$ 
let us set in (\ref{App_Phi2_Int}) the terms standing at $R^{-1}$ 
and at all $t^{2k}$ equal to zero. 
Making use of (\ref{App_R2elim}) this results in restriction 
for expansion coefficients 
(assuming $\Phi_j$ is normalized by setting $C_{0,1,0}=1$) 
\begin{eqnarray}
C_{1,1,1}&=&
\frac{5}{24}\zeta-\frac{5}{24}C_{1,1,0}-\frac{\sqrt{2} Z}{6}
\label{App_C111}
\\
C_{1,1,k+1}&=&
\frac{1}{(2k+3)(2k+2)} \Big [\big (2k(2k+2)-\frac{5}{4} \big )C_{1,1,k}-\frac{\sqrt{2} Z (4k-1)!!}{(2k+1)(4k)!!} \Big ], 
\qquad k=1,2,\dots
\nonumber
\\
\label{App_C11k}
\end{eqnarray}
Thus $\Phi_1$ ,$\Phi_2$ and of course $\Phi_0$ defined by (\ref{App_Phi0}) can be correctly introduced without any need to incorporate negative powers of $R$.
Using the convergence criterion for power series 
it can be shown that the function
$f(t)=\sum_{k=0}^{\infty}C_{1,1,k}t^{2k}$ defined by the sequence 
(\ref{App_C11k}) converges uniformly to a finite function on its 
whole interval of definition ($t \in (-1,1)$).

Let us analyse $\Phi_{j}$ for the case $j>2$ now. 
First, let us notice that if the lowest power of $R$ in $\Phi_j$ is $R^i$ 
($i\neq 0$) 
then the effect of acting $H_j$ on $\Phi_j$ 
(for the case when (\ref{App_R2elim}) holds) 
is lowering the lowest power or $R$ to $R^{i-1}$. 
In accord with this, 
as $\Phi_1$, $\Phi_2$ we alredy managed to construct singularity-free, 
the only singularity in $\Phi_3$ (of the type $R^{-1}$)  
can come from acting $H_1$ on $\Phi_1$. 
The only possibility to eliminate this singularity is now to allow 
the same singularity $R^{-1}$ to occur in $\Phi_3$. 
$\Phi_4$ must contain the $R^{-1}$ term again 
(as both $\Phi_3$ and $H_2 \Phi_2$ contain it) and also the $R^{-2}$ 
due to the fact, that 
now a non-zero term with $R^0$ also occurs.

Generally, notice that while the effect of acting of (\ref{App_Hj}) 
on a term containing no $R$-type singularities ($\Phi_1$, $\Phi_2$) 
was the occurence of $R^{-1}$ singularity only,  
each acting of (\ref{App_Hj}) on $R^{-i}, i>0$ produces also an $R^{-(i+2)}$ 
singular term.
Consequently, for each $j>2$, if $\Phi_j$ contains the $R^{-i}$ term, 
$\Phi_{j+2}$ must contain $R^{-(i+2)}$ 
[plus the singularity coming from $\Phi_{j+1}$ which can, however, 
never be worse than $R^{-(i+2)}$]. 
Starting from the fact that $\Phi_3$ contains $R^{-1}$, 
$\Phi_4$ contains $R^{-1}$ and $R^{-2}$, etc.,
it is now easy to show recursively that $\Phi_{j+2}$, $j=1,2,\dots$ must contain the $R^{-j}$ term, 
that is, terms $R^{-i}$ must be incorporated into (\ref{App_PhiJ}) with  
$i_{max}=j-2$, for $j>2$. In this way, all the $R^{-1}$, $R^{-2}$ singularities have a chance to be eliminated.

Up till now we only showed, that the $R^{-i}$ singularities 
can be eliminated from the set of equations (\ref{App_Phi}).
By construction, the singularities are evidently still present 
in the functions $\Phi_j$ and, apparently, 
also in the wave function (\ref{App_Psi}). 
One has, however, to realise that 
${r_{12}}/{R}$
is always finite [see (2)-(5)], as
\begin{eqnarray}
\frac{r_{12}}{R} \leq 2^{1/2}.
\label{App_r12Rlim}
\end{eqnarray}
For our construction of $\Phi$'s the worst combination of $R$ and $r_{12}$ 
will evidently be 
${r_{12}^{j}}/{R^{j-2}}$, 
which is, according to (\ref{App_r12Rlim}) 
always finite, as well.

Let us analyze what happens when the infinite series (\ref{App_Psi})  
is truncated (which is always the case in practice). 
As an example let us consider the case when in (\ref{App_PhiJ}) 
$i_{max}=1$ a $j$ is finite 
In the following we will analyze the most problematic series 
resulting from (\ref{App_PhiRec}).
Let us introduce
\begin{eqnarray}
D_{j,k}=\frac{(2)^J}{(j+3+4k)(j+1+4k)...(j_{0}+4k)(2(j+2))!!},
\end{eqnarray}
where $J=(j+2)/2$ and $j_{0}=3$ for even $j$ 
and $J=(j+1)/2$ and $j_{0}=4$ for odd $j$.  
Acting repeatedly with the 
$-\frac{\partial^{2}}{\partial R^{2}}$ part of $H_j$ on $\Phi_j$ yields 
\begin{eqnarray}
\sum_{k=0}^{\infty}C_{1,1,k}D_{j,k}(\frac{1}{2}\zeta^{2}+E)^{J}
Rt^{2k}\exp{-\zeta R},
\label{App_Rad1}
\end{eqnarray}
The dominant series resulting from recursive acting of 
$-\frac{2(1-t^{2})}{R^{2}}\frac{\partial^{2}}{\partial t^{2}}+\frac{6t}{R^{2}}\frac{\partial}{\partial t}$ 
is 
\begin{eqnarray}
\sum_{k=1}^{\infty}D_{j,k}C_{1,1,k}\big(\frac{16k^{2}-\frac{11k}{2}+\frac{5}{4}}{2k+3}\big )^{J}
R^{1-2J}t^{2k}\exp{-\zeta R},
\label{App_Rad2}
\end{eqnarray}
Finally, acting of 
$\frac{2jt}{R^{2}}\frac{\partial}{\partial t}$ 
on $\Phi_j$ returns
\begin{eqnarray}
\sum_{k=0}^{\infty}C_{1,1,k}D_{j,k}(2j)!!(2k)^{J}R^{1-2J}t^{2k}\exp{-\zeta R}.
\label{App_Rad3}
\end{eqnarray}
All the other parts of $H_j \Phi_j$ produce series which are minorant 
to one of series (\ref{App_Rad1}-\ref{App_Rad3}).  
Thus, they need not be analysed. 
Let us consider the convergence of (\ref{App_C11k}) 
and the restriction (\ref{App_r12Rlim}) now. 
Evidently, the series (\ref{App_Rad1}) converges uniformly when 
$\zeta^{2}/{2} + E < 1$. 
This statement was confirmed also by numerical results. 
The series (\ref{App_Rad2}) uniformly converges for 
 $-1 < t < 1$.
Similarly, the series (\ref{App_Rad3}) converges for finite $j$. 
As all the other series are minorant to some of the cases 
(\ref{App_Rad1}-\ref{App_Rad3}),  
also the complete sum of series converges to a finite function for a finite $j$.

%

\section{Acknowledgement}
The author is deeply indebted to Pavel Neogr\'{a}dy and J\'{a}n Ma\v{s}ek for many 
inspirating discussions and to Ilja Marti\v{s}ovits for discussions of mathematical aspect of the
problems. Special thanks are due to \v{S}tefan Varga for help with 
calculation of some integrals and for help with preparing of the manuscript and
also to \v{S}tefan Dobi\v{s} for technical support.
This work is dedicated to the memory of Ladislav Turi-Nagy the former head of Department 
of Theoretical Chemistry at the Institute of Inorganic Chemistry of Slovak Academy of Sciences,
a man who encouraged me very much in my work.

\end{document}